\documentclass{article}

\usepackage{arxiv}

\usepackage[utf8]{inputenc} % allow utf-8 input
\usepackage[T1]{fontenc}    % use 8-bit T1 fonts
\usepackage{hyperref}       % hyperlinks
\usepackage{url}            % simple URL typesetting
\usepackage{booktabs}       % professional-quality tables
\usepackage{amsfonts}       % blackboard math symbols
\usepackage{nicefrac}       % compact symbols for 1/2, etc.
\usepackage{microtype}      % microtypography
\usepackage{cleveref}       % smart cross-referencing
\usepackage{lipsum}         % Can be removed after putting your text content
\usepackage{graphicx}
\usepackage{doi}

\title{Automated brain parcellation rendering and visualization in R with \textit{coldcuts}}

\author{ \href{https://orcid.org/0000-0001-8315-7406}{\includegraphics[scale=0.06]{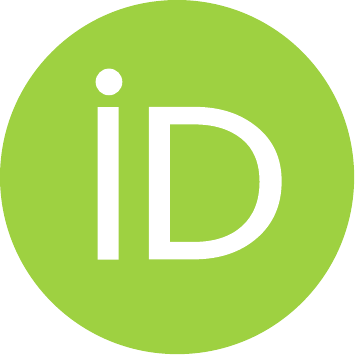}\hspace{1mm}Giuseppe A.~D'Agostino}\\
	Lee Kong Chian School of Medicine\\
	Nanyang Technological University\\
	Singapore, Singapore 308207 \\
	\texttt{giuseppe.ad@ntu.edu.sg} \\
	\And
	\href{https://orcid.org/0000-0003-4419-476X}{\includegraphics[scale=0.06]{orcid.pdf}\hspace{1mm}Sarah R.~Langley} \\
	Lee Kong Chian School of Medicine\\
	Nanyang Technological University\\
	Singapore, Singapore 308207 \\
	\texttt{sarah.langley@ntu.edu.sg} \\
}

\renewcommand{\headeright}{\textit{arXiv} }
\renewcommand{\undertitle}{\textit{arXiv} }
\begin{document}
\maketitle

\begin{abstract}
Parcellations are fundamental tools in neuroanatomy, allowing researchers to place functional imaging and molecular data within a structural context in the brain. Visualizing these parcellations is critical to guide biological understanding of clinical and experimental datasets in humans and model organisms. However, software used to visualize parcellations is different from the one used to analyze these datasets, greatly limiting the visualization of experimental data within parcellations. We present \textit{coldcuts}, an open source R package that allows to automatically generate, store and visualize any volume-based parcellation easily and with minimal manual curation. \textit{coldcuts} allows to integrate external datasets and offers rich 2D and 3D visualizations. \textit{coldcuts} is freely available at \href{http://github.com/langleylab/coldcuts}{http://github.com/langleylab/coldcuts} and several curated \textit{coldcuts} objects are made available for human, mouse, chimpanzee and \textit{Drosophila} parcellations at \href{https://github.com/langleylab/coldcuts_segmentations}{https://github.com/langleylab/coldcuts\_segmentations}.
\end{abstract}

% keywords can be removed
\keywords{neuroanatomy \and parcellations  \and dataviz }

\section*{Introduction}
With the ever increasing amount of neurological data being generated from imaging techniques and molecular profiling, there is an ongoing need for data visualizations that generate, integrate and display anatomical information in a concise and informative way.  This demand is not limited to human brain visualization but for model organisms as well, given that both fundamental and translational research utilize models such as mouse, non-human primates, and \textit{Drosophila}.  With imaging technologies such as Magnetic Resonance Imaging, Computed Tomography and lightsheet microscopy, neuroanatomists have been able to create detailed descriptions of regions and structures in animal brains. By accumulating several neuroimaging datasets and applying computational geometry algorithms, reference spaces are established for each species, in which the position, size and shape of each structure is standardized \cite{Eickhoff2018-wi}. Anatomists then divide these standardized spaces in relevant non-overlapping structures, thus creating a brain parcellation (also known as segmentation). These parcellations can be visualized by stand-alone tools, which allow their rendering in 2D and 3D, their overlap with other neuroimaging samples and, in some cases, their modification. 

There are, however, other modalities in which brain-derived data is processed in more flexible statistical programming environments (such as the R computing language), for instance in the case of tissue-resolved transcriptomics. A crosstalk between the results of an analysis in R and the tools commonly used in segmentation visualization is not trivial and requires additional coding in different programming languages. There are several packages within the R language which create high quality representations of brain parcellations \cite{Bahl2017-um,Schafer2020.09.18.302935,Mowinckel2020-bu,10.7554_eLife.53350} but they are currently limited by their focus on single species brain and, in some cases, their dependence on manual curation. Here, we present \textit{coldcuts}, an R package for the automatic rendering and plotting of any kind of volume-based parcellation, regardless of species, and for its integration with external data.

\section*{Implementation}

\textit{coldcuts} reads several formats of 3D parcellation in R as 3D arrays, slices along the three anatomical planes, and calculates structural contours for every structure in every slice separately. Contour calculation is achieved by applying a fast marching squares algorithm as implemented in the \textit{isoband} package \cite{Wilke2021-md}. These contours are then rendered as polygons using \textit{ggplot2} \cite{Wickham2009-mi}, calculating which contours constitute holes, and which are instead filled. Parcellations can then be plotted separately for every anatomical plane selecting a specific slice (fig. 1A) (corresponding to a 1-voxel thick slice in the Left-Right, Anterior-Posterior, or Inferior-Superior direction), with colours assigned to each structure according to a structural ontology provided by the user (fig. 1B). Optionally, the user can overlay structure acronym labels on the plot (fig. 1A), and/or subset the parcellation to visualize only specific structures. If the user wants to visualize structures as a whole rather than on single slices, e.g. to see structures that occur on different parts of the brain, coldcuts can compute a maximum projection of every structure on both sides of an anatomical axis (fig. 1C). Slices, ontology, metadata and projections are all stored in the S4 \textit{segmentation} class, which also hosts \textit{assays}. \textit{Assays} contain numerical data for several features (such as genes) to be plotted in a parcellation, e.g. bulk transcriptomic data measured in different brain areas. For an assay to be represented, the user must specify a mapping between the assay columns - the samples - and the structures. 

\textit{Segmentation} class objects are meant to be shared for easy access and visualization, and their assay slot can be easily updated with new data to be represented in that specific parcellation. A key component of a \textit{segmentation} object is its \textit{ontology}, a table inspired by the Allen Human Reference Atlas structural ontology \cite{Wickham2009-mi,Ding2016-lc}, which describes each structure in terms of name, acronym, colour and hierarchical relationships to other structures (fig. 1B). For instance, the “thalamus” group can be divided into several thalamic nuclei, each of which is considered as a separate volume in the Allen Human Reference Atlas. Users can subset and plot structure groups by indicating a more upstream parent node.  \textit{coldcuts} simplifies polygons as much as possible by storing essential information and smoothing polygons in draw time only using kernel smoothing \cite{RSMOOTH}. However, for parcellations at high resolution, \textit{coldcuts} objects can be quite large, occupying several GB of RAM in R (e.g. 1.9 GB for $7.7 \times 10^{7}$ voxels, 7 GB for $1.2 \times 10^{9}$ voxels) . For this reason, \textit{coldcuts} lets the user select one specific anatomical plane, or subset segmentations by slice and/or structure, in order to allow sharing lightweight versions of the segmentation which include only the areas or structures of interest.  Finally, \textit{coldcuts} uses marching cubes, quadric edge decimation and Laplacian smoothing (as implemented in the \textit{Rvcg} package \cite{Schlager2017nz}) to render lightweight 3D meshes of the parcellations, which the user can interactively visualize in R using \textit{rgl} \cite{Murdoch2021ym} (Fig 1D).

%\label{sec:headings}

%\lipsum[4] See Section \ref{sec:headings}.
%
%\subsection{Headings: second level}
%\lipsum[5]
%\begin{equation}
%	\xi _{ij}(t)=P(x_{t}=i,x_{t+1}=j|y,v,w;\theta)= {\frac {\alpha _{i}(t)a^{w_t}_{ij}\beta _{j}(t+1)b^{v_{t+1}}_{j}(y_{t+1})}{\sum _{i=1}^{N} \sum _{j=1}^{N} \alpha _{i}(t)a^{w_t}_{ij}\beta _{j}(t+1)b^{v_{t+1}}_{j}(y_{t+1})}}
%\end{equation}
%
%\subsubsection{Headings: third level}
%\lipsum[6]
%
%\paragraph{Paragraph}
%\lipsum[7]
%

\section*{Application}
As an example, we show plots for the \textit{coldcuts} segmentations of the Allen Human Reference Atlas (figure 1A, 1C), the Janelia Farms Drosophila Brain Atlas JFC2010 \cite{Ito2014-bu,Bogovic2020-uz}  (figure 1E), and the Allen Mouse Brain Atlas CCFv3 \cite{Lein2007-fw,Wang2020-sb} (figure 1F). To demonstrate the assay visualization functionality in the context of a human parcellation, we show the results of plotting an assay containing GTEx RNA-seq normalized gene expression values \cite{GTEx_Consortium2013-jn} within the maximum projection of the structures that were included in each dataset (figure 1G) for the beta-amyloid precursor peptide (\textit{APP}) gene, which is causally linked to early-onset Alzheimer’s disease \cite{Goate2006-oe}. This visualization easily shows that, in GTEx samples - obtained by individuals that were neurologically healthy - the cingulate gyrus is particularly affected, followed by hypothalamic nuclei and the hippocampus. Atrophy of the cingulate gyrus has been involved in early phases of disease onset \cite{Scahill2002-qa}, and this visualization corroborates this evidence in a simple and intuitive way. Several segmentation class objects have been generated from publicly available volume-based parcellations, together with curated ontologies and metadata, and they can be found at \href{https://github.com/langleylab/coldcuts\_segmentations}{https://github.com/langleylab/coldcuts\_segmentations}.

\section*{Conclusion}
We believe \textit{coldcuts} to be an easy to use and useful tool for the neuroimaging and neurogenomics community, especially when different data modalities need to be integrated in the framework of a parcellation, or several parcellations need to be compared within the R programming language. In the absence of manually curated parcellation renderings for other species, \textit{coldcuts} allows to easily fill in the gap and provide fast and simple access to these representations in R. \textit{coldcuts} is available as an R package on GitHub at \href{http://github.com/langleylab/coldcuts}{http://github.com/langleylab/coldcuts}, together with curated segmentation objects for human, chimp, mouse and \textit{Drosophila} brains \cite{Glasser2016-my,Hammers2003-ev,Faillenot2017-lw,Vickery2020-ph,Ding2016-lc,Lein2007-fw,Wang2020-sb,Ito2014-bu,Bogovic2020-uz}.

\begin{figure}
	\centering
	\includegraphics[scale=0.65]{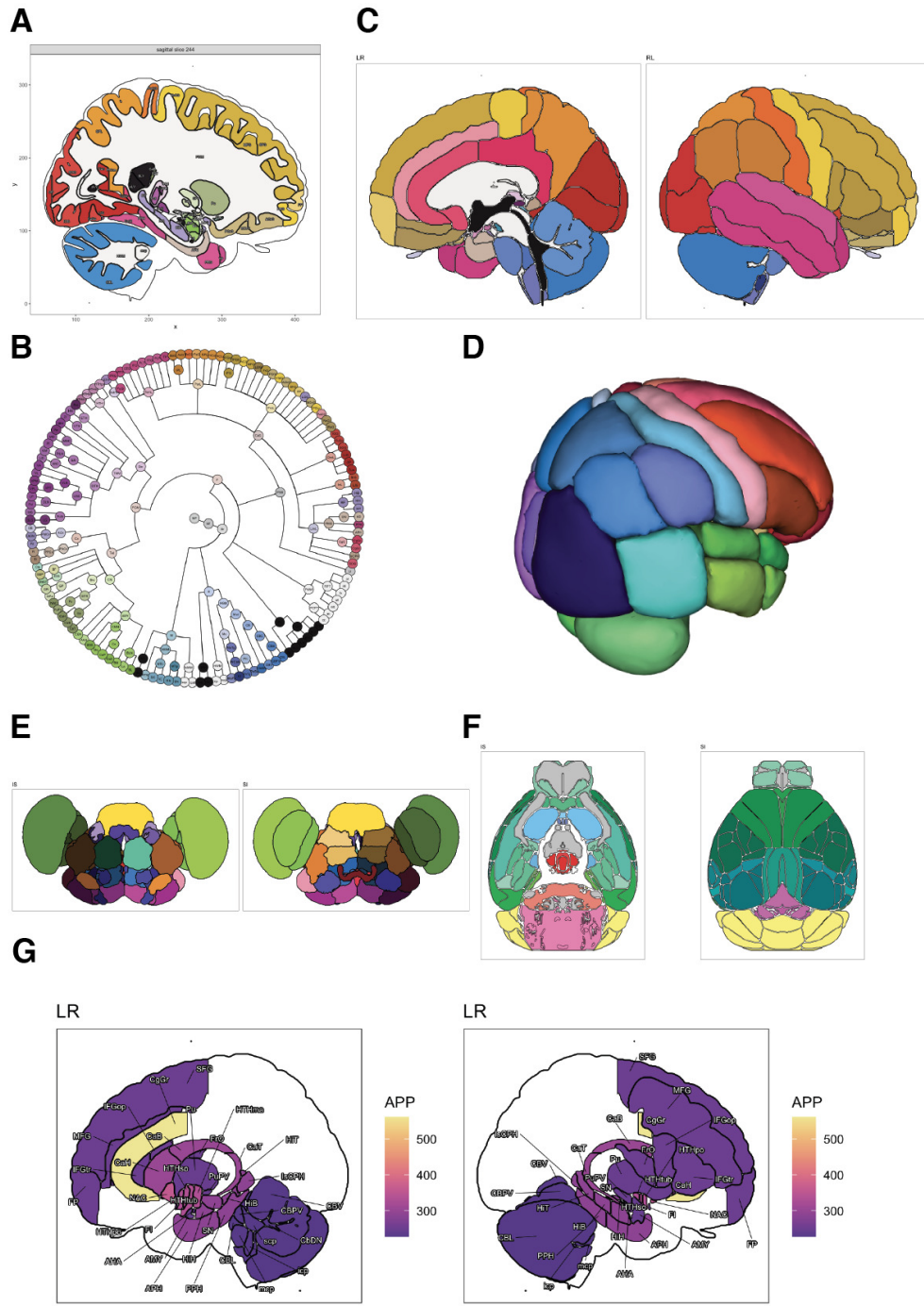}
	\caption{Visualizations created with the \textit{coldcuts} package. \textbf{A:} visualization of one sagittal slice from the Allen Institute for Brain Sciences (AIBS) human brain segmentation, including structure labels; \textbf{B:} structural ontology graph for the AIBS human brain segmentation; \textbf{C:} maximum projection on the sagittal plane from the left and right viewpoints; \textbf{D:} 3D mesh visualization of the Hammersmith human brain atlas segmentation; \textbf{E:} maximum projection on the axial plane from inferior and superior viewpoints of the JRC2010 \textit{Drosophila melanogaster} segmentation; \textbf{F:} maximum projection on the axial plane from inferior and superior viewpoints of the CCFv3 \textit{Mus musculus} segmentation; \textbf{G:} visualization of median Transcripts Per Million (TPM) for the \textit{APP} gene from the GTEx RNA-seq data on maximum projection of the sagittal plane for a subset of sampled structures.}
	\label{fig:fig1}
\end{figure}

\section*{Acknowledgments}
GAD conceived and implemented the software, curated segmentation objects, wrote the manuscript. SRL coordinated the work, obtained funding, wrote the manuscript. We thank members of the Integrative Biology of Disease Lab at Lee Kong Chian School of Medicine, Singapore, for testing the software. SRL is supported by the Lee Kong Chian School of Medicine and Nanyang Technological University Singapore Nanyang Assistant Professor Start-Up Grant. GAD is supported by the Lee Kong Chian School of Medicine Dean's Postdoctoral Fellowship.

%\label{sec:others}
%
%\subsection{Citations}
%Citations use \verb+natbib+. The documentation may be found at
%\begin{center}
%	\url{http://mirrors.ctan.org/macros/latex/contrib/natbib/natnotes.pdf}
%\end{center}
%
%Here is an example usage of the two main commands (\verb+citet+ and \verb+citep+): Some people thought a thing \citep{kour2014real, hadash2018estimate} but other people thought something else \citep{kour2014fast}. Many people have speculated that if we knew exactly why \citet{kour2014fast} thought this\dots
%
%\subsection{Figures}
%\lipsum[10]
%See Figure \ref{fig:fig1}. Here is how you add footnotes. \footnote{Sample of the first footnote.}
%\lipsum[11]
%
%\begin{figure}
%	\centering
%	\fbox{\rule[-.5cm]{4cm}{4cm} \rule[-.5cm]{4cm}{0cm}}
%	\caption{Sample figure caption.}
%	\label{fig:fig1}
%\end{figure}
%
%\subsection{Tables}
%See awesome Table~\ref{tab:table}.
%
%The documentation for \verb+booktabs+ (`Publication quality tables in LaTeX') is available from:
%\begin{center}
%	\url{https://www.ctan.org/pkg/booktabs}
%\end{center}
%
%
%\begin{table}
%	\caption{Sample table title}
%	\centering
%	\begin{tabular}{lll}
%		\toprule
%		\multicolumn{2}{c}{Part}                   \\
%		\cmidrule(r){1-2}
%		Name     & Description     & Size ($\mu$m) \\
%		\midrule
%		Dendrite & Input terminal  & $\sim$100     \\
%		Axon     & Output terminal & $\sim$10      \\
%		Soma     & Cell body       & up to $10^6$  \\
%		\bottomrule
%	\end{tabular}
%	\label{tab:table}
%\end{table}
%
%\subsection{Lists}
%\begin{itemize}
%	\item Lorem ipsum dolor sit amet
%	\item consectetur adipiscing elit.
%	\item Aliquam dignissim blandit est, in dictum tortor gravida eget. In ac rutrum magna.
%\end{itemize}

\bibliographystyle{ieeetr}
%\bibliography{references}  %%% Uncomment this line and comment out the ``thebibliography'' section below to use the external .bib file (using bibtex) .


%

\end{document}